\documentclass[mathleft]{an}
\usepackage{graphicx}
\usepackage{times}
\begin{document}

\Pagespan{789}{}
\Yearpublication{2006}%
\Yearsubmission{2005}%
\Month{11}%
\Volume{999}%
\Issue{88}%

\title{Modelling $\delta$ Scuti stars using asteroseismic space data}

\author{A. Moya\inst{1}\fnmsep\thanks{Corresponding author:
  \email{amoya@cab.inta-csic.es}\newline}, A. Garc\'ia Hern\'andez\inst{2}, J.-C. Su{\'a}rez\inst{2}, C. Rodr\'iguez-L\'opez\inst{3,4} \and R. Garrido\inst{2}}

\titlerunning{Modelling $\delta$ Scuti stars using asteroseismic space data} 
\authorrunning{A. Moya et al.}
\institute{ 
Departamento de Astrof\'{\i}sica, Laboratorio de
  Astrof\'{\i}sica Estelar y Exoplanetas, LAEX-CAB (INTA-CSIC), PO BOX
  78, 28691 Villanueva de la Ca\~nada, Madrid, Spain 
\and 
Instituto de
  Astrof\'isica de Andaluc\'ia, IAA - CSIC, Granada, Spain E-18008
\and 
Laboratoire d'Astrophysique de Toulouse-Tarbes, CNRS,
  Universit\'e de Toulouse. Toulouse, France F-31400
\and 
Departamento de F\'\i sica Aplicada. Universidad de Vigo, Vigo, Spain E-36310
}
\received{30 May 2005}
\accepted{-}
\publonline{later}


\abstract{In the last years, space missions such as COROT, Kepler or
  MOST have provided very accurate photometric observational data. In
  the particular case of $\delta$ Scuti stars, the observed frequency
  spectra have hundreds (if not thousands) of modes and a clear
  amplitude distribution. In this work we present new techniques for
  modelling these observations and the results obtained. We searched
  for regular patterns in the observational data, which yields
  something resembling the large separation. This allows to reduce the
  possible positions of the star in the HR diagram, yielding a value
  of the mean density with an accuracy never reached before for
  isolated stars of this type. Finally, we answer whether a $\delta$
  Scuti star is stable despite all of the observed frequencies are
  simultaneously excited}

\maketitle

\section{Introduction\label{sec:intro}}


The $\delta$ Scuti stars are intermediate-mass pulsating variables
with spectral types ranging from A2 to F0. They are located on and
just off the main sequence in the lower part of the Cepheid
instability strip (luminosity classes V \& IV). Nowadays, the $\delta$
Scuti stars are considered as particularly suitable for asteroseismic
studies of poorly known hydrodynamical processes occurring in stellar
interiors such as convective overshoot, mixing of chemical elements
and redistribution of angular momentum (Zahn 1992), etc. Due to the
complexity of the oscillation spectra, their pulsating behaviour is
not fully understood, in particular in what regards the
rotation-pulsation interaction (see a complete rewiew on such effects
in Goupil et al. 2005). In the last decade, numerous interpretation
works have taken the effects of rotation into account (Foz Machado et
al. 2006; Su\'arez et al. 2007a; Su\'arez et al. 2007b; Bruntt et
al. 2007)


The very precise space photometry supplied by the CoRoT mission give
us the possibility to deal with a range and an amount of frequencies
not reached by usual ground based observations. In this work we focus
on the star HD\-174936. This is a field $\delta$ Scuti star observed
during the first short run SRc01 for which we have a frequency
resolution corresponding to 27 days of observation, namely
0.45~$\mu$Hz.

The evolutionary code CESAM (Morel 1997; Morel \& Lebreton 2008) and
the pulsation codes GraCo (Moya et al. 2004; Moya \& Garrido 2008)
have been used as numerical codes to calculate frequencies, growth
rates and other physical quantities. GraCo provides, in particular,
non-adiabatic variables and growth rates. Comparing the theoretical
predictions given by these numerical codes with the observed range of
excited frequencies and studying the periodic spacing in the
oscillation frequency distribution, we have performed a seismic study
in order to constrain the physical parameters of the star.

Finally, we have used a representative model of HD\-174936 to study if
the star can excite, at the same time, the huge number of pulsational
modes observed in $\delta$ Scuti stars from space (Poretti et
al. 2009; Moya \& Rodr\'iguez-L\'opez 2010).

\section{Analysis of the data \label{sec:data}}

The star was observed during the first short run of CoRoT. Around 27
days of very precise photometry at a sampling time of 32~s is the data
base used for the frequency analysis presented here. The spectral
window of the satellite observations is almost perfect (less than 5\%
of side lobes). A detailed explanation of the analysis of this star
can be found in Garc\'ia Hern\'andez (2009).



\begin{figure}
 \begin{center}
  \scalebox{.20}{\includegraphics{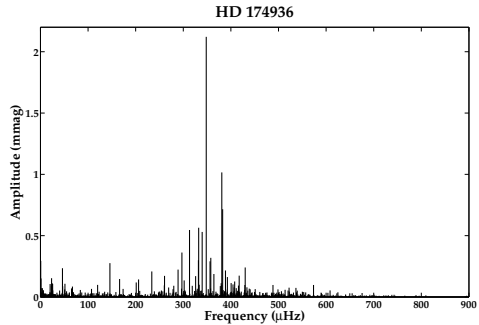}}
  \caption{A total number of 422 frequencies, ranging from 50 to
    1000~$\mu$Hz, have been obtained. The highest peaks on amplitude
    fall in the interval of 250 to 450 $\mu$Hz.}
  \label{fig:spectrum}
 \end{center}
\end{figure}

The time series was analysed using the prescriptions given in Reegen
(2007) and Lenz \& Breger (2005). If we adopt the rather conservative
criteria to stop the search for frequencies when the significance
level is 10 (see Reegen 2007), then 422 frequencies are found. The
obtained power spectrum is shown in Fig.~\ref{fig:spectrum}. More
peaks could be extracted because of the high significance of the
corresponding fitting, but the prewhitening procedure becomes
exponential and we can not be sure whether we are filtering real
frequencies from the object or any other artifact due to the
instrument or intrinsic to the mathematical technique (Poretti et
al. 2009).

Recently , Kallinger \& Matthews (2010) have suggested that the lowest
amplitude modes observed in $\delta$ Scuti stars from space can be due
to surface granulation, and as a consequence only some tenths of modes
with the highest amplitude would be real pulsation modes.

Mode identification, mainly for $\delta$ Scuti stars, is one of the
bottle necks for the progress in asteroseismology. Individual mode
identification using white light, as supplied by the CoRoT photometer,
is not possible. However, statistical properties of the full stellar
oscillation spectrum provide us with information related with global
properties of the stellar model. In particular, the classical large
($\Delta\nu=\nu_{n+1,\ell}-\nu_{n,\ell}$) and small
($\delta\nu=\nu_{n,\ell}-\nu_{n-1,\ell+2}$) separations, frequently
used in the solar context, could also be applied to this
star. However, some caution has to be taken, because for $\delta$
Scuti main sequence stars, the excited oscillation modes are not
expected to be in the asymptotic regime.

In order to search for periodic patterns in the frequency set, we
consider them as a set of Dirac's $\delta$'s of equal amplitude,
centred at each frequency value. If the frequencies were equally
spaced, then we will have a Dirac comb with Fourier transform being
another Dirac comb with periodic inverse values and multiples of the
original spacing. That means that if we have a given periodicity in
the frequency set of $\delta f$, then we will have a peak in the
Fourier Transform of $\delta f$ and its sub-multiples, i.e. $\delta
f/n$ with $n=1, 2, 3...$. Although Handler et al. (2000) uses a
similar technique, they do not give any details of the method.

\begin{figure*}
 \begin{center}
  \scalebox{.20}{\includegraphics{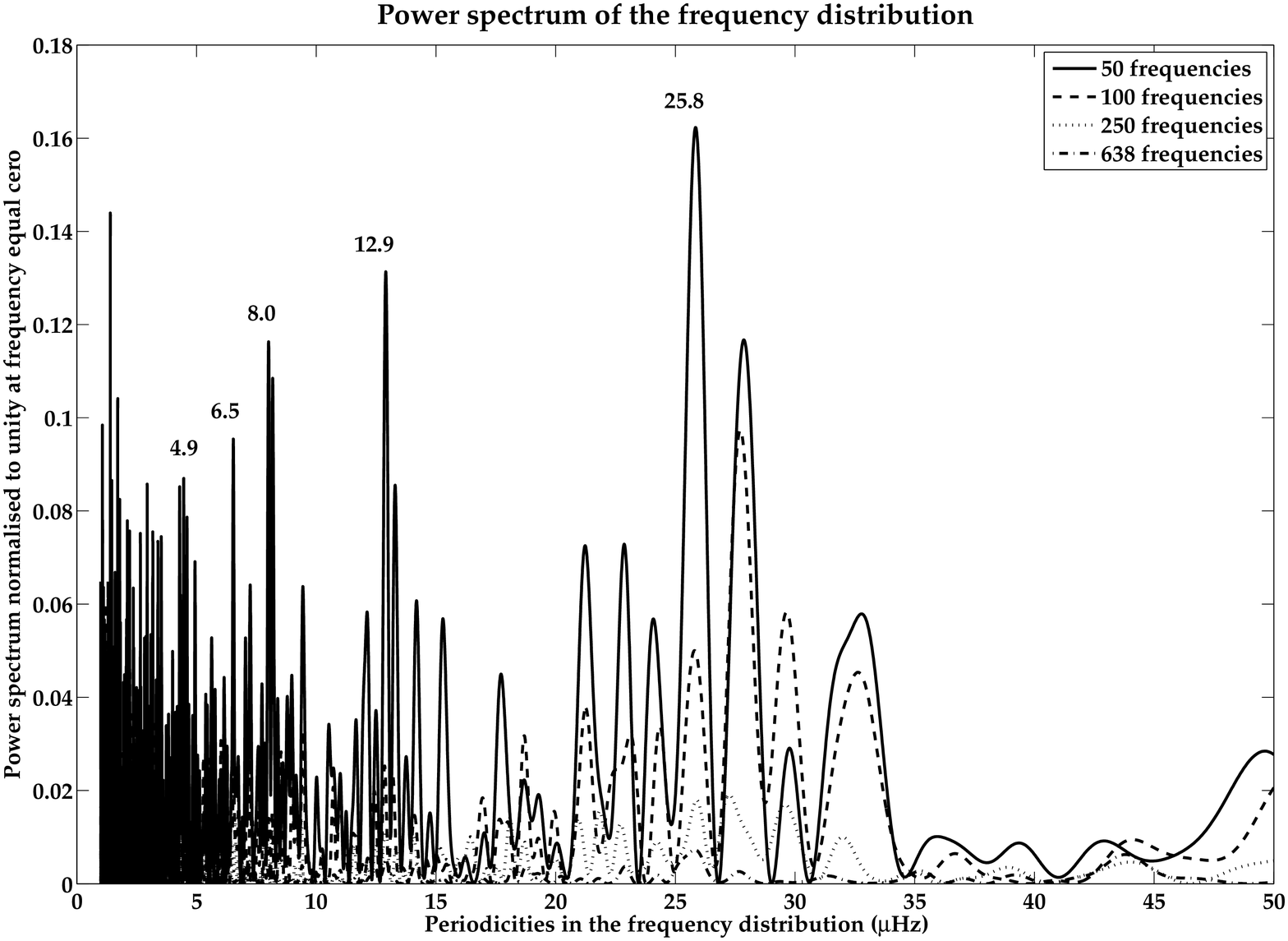}}
  \scalebox{.31}{\includegraphics{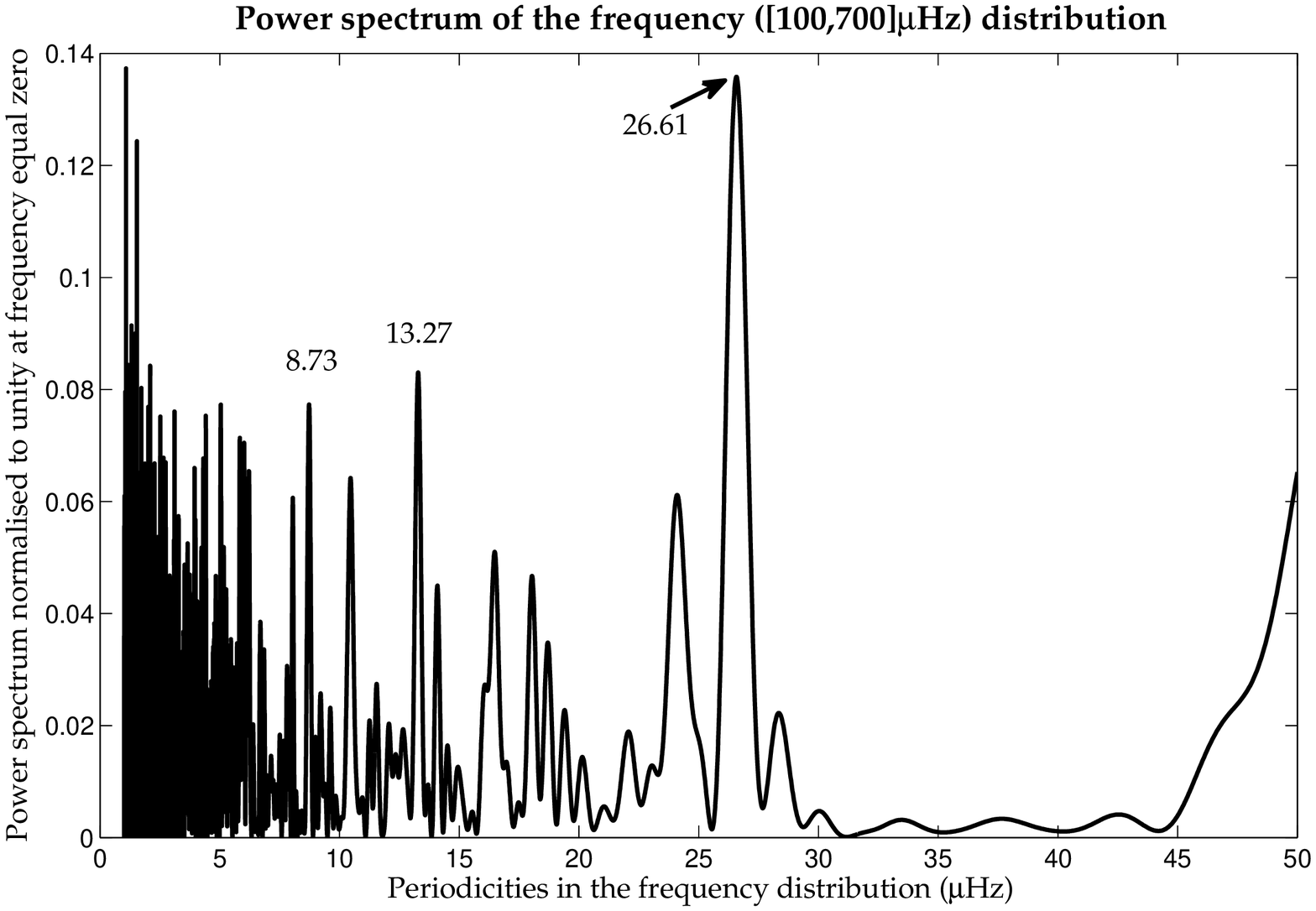}}
  \caption{Left panel: Power spectrum for various subsets of
    frequencies, selected by amplitudes. The solid line represents the
    power spectrum when the highest 50 frequencies are selected; the
    dashed line correspond to the highest 100 frequencies; the dotted
    line to 250; and the dot-dashed line to all of them. It can be
    seen how the peaks disappear as the number of frequencies
    increases. The peak corresponding to the large separation (25.8
    $\mu\mbox{Hz}$) and its sub-multiples are labelled. Right panel:
    Power spectrum of the theoretical frequencies calculated for a
    representative model of HD\,174936. Only modes with frequencies in
    the range [100,700] and with $\ell=0-3$ have been taken into
    account. The peak at 26.61~$\mu$Hz (and sub-multiples) is the half
    of the possible large separation obtained using these modes.}
  \label{fig:large-sep}
 \end{center}
\end{figure*}

We started this analysis assuming the hypothesis that, selecting a
subset with the highest peaks of the entire frequency set, we are
actually considering mainly the modes of the lowest $\ell$ values. The
visibility of the modes decreases approximately as $\ell^{-2.5}$ or
$\ell^{-3.5}$, depending whether the $\ell$ degree is odd or even
(Dziembowski 1977). Although some authors have identified higher
$\ell$ values in the frequency spectrum of a star observed as a whole
(Daszy\'nska-Daszkiewicz et al. 2006), in general, the selection of
the highest amplitude modes ensures that most of the modes have
$\ell<4$, enough to obtain general properties with an statistical study.

Therefore, we use the hypothesis that, selecting a subset of
frequencies with the highest amplitudes, we select mainly the
frequencies of lowest $\ell$ values. To probe that, we have
constructed several subsets with an arbitrary number of frequencies
taking always the highest peaks and applied the Fourier transform to
them. We have applied the method to HD\,174936:
Fig. \ref{fig:large-sep} (left panel) shows that the greater the
number of frequencies in the subset, the lower the peaks. Clearly,
when we select 50 frequencies a periodic pattern can be recognised,
because of the simultaneous appearance of peaks at 25.8, 13, 8, 6.5
and 4.9~$\mu$Hz. This indicates that we probably have a Dirac comb of
25.8~$\mu$Hz, which is the signature of a large separation of around
52~$\mu$Hz.  We adopted a range of values of [45,60] $\mu$Hz to be
representative of this large separation. Other peaks surrounding the
main one at 25.8 $\mu$Hz (and sub-multiples) may be produced by the
quasi-periodic ``sampling'' exhibited by the frequency set. The rest of
shorter periodicities are less clear to explain, because the periodic
pattern of small separations and g-modes are of the same order.

\section{The modelling \label{sec:models}}

The stellar models were computed with the evolutionary code CESAM
following the recommendations of the ESTA group studies (Lebreton et
al. 2008; Moya et al. 2008). The physics used is that suitable for this
stellar type, and used in numerous studies (Casas et al. 2006; Casas
et al. 2009; Garc\'ia Hern\'andez et al. 2009).

Theoretical oscillation spectra were computed for the equilibrium
models using the GRAnada oscillation COde provides diagnostics on the
instability and non-adiabatic observables. In this code the
non-adiabatic pulsation equations are solved following Unno et
al. (1989).

\section{Constraining the seismic models of HD\,174936 \label{sec:results}}


\begin{figure}
 \begin{center}
  \scalebox{.20}{\includegraphics{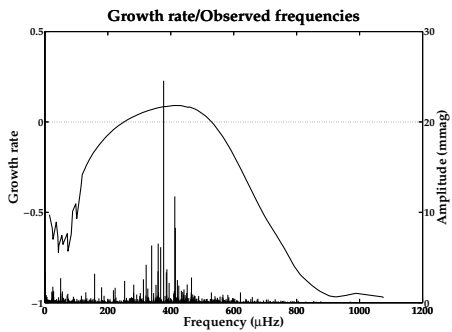}}
  \caption{Frequency versus growth rate and amplitude. Growth rate
    (solid line, left axis) shows the stability range ($<$ 0) and one
    vertical line of the amplitude height for each frequency is
    represented (right axis).}
  \label{fig:growth}
 \end{center}
\end{figure}

HD\,174936 has a $T_{\rm eff}=8000\pm 200$K, a $\log g=4.08\pm 0.2$
and a $[Fe/H]=-0.32\pm0.2$ . These values are taken from ``CorotSky
Database'' (Charpinet et al. 2006). The rotational velocity ($v\,\sin
i = 169.7$ km/s) has been determined from high-resolution
spectroscopy. It was obtained in the framework of the mission
preparation and it is available at the GAUDI archive (Solano et
al. 2005). It has been shown that rapid rotation should be considered
when calculating the stellar physical parameters from
photometry. Following Michel et al. (1998); Su\'arez et al. (2002),
and considering the absence of additional information on the
inclination angle of the star, uncertainties in the HR-Diagram box of
200~K in $T_{\rm eff}$ and $\sim$ 0.2~dex in $logg$ are adopted.

\begin{figure}
 \begin{center}
  \scalebox{.26}{\includegraphics{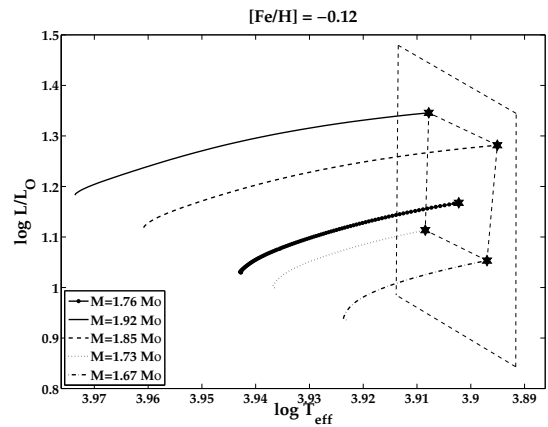}}
  \caption{HR diagram showing the observational photometric
    uncertainty box (the large one) for HD\,174936. The small box is
    that obtained when the asteroseismological constraints are
    used. Filled stars correspond to models representative of the
    star.}
  \label{fig:box1}
 \end{center}
\end{figure}

We constructed models in the HR diagram corresponding to the four
corners and the center of the uncertainty box, with the observed
metallicity and its errors (see Fig.~\ref{fig:box1}), $\alpha_{\rm
  MLT}=0.5$ and overshooting 0.2. Then, for each equilibrium model we
computed the corresponding adiabatic and non-adiabatic oscillations,
and from the whole set of models, we selected those fitting the
average large separation defined in Sect. 2. Finally, from the
remaining subset of models we selected those predicting unstable the
observed frequencies with the largest amplitudes. The properties of
the resulting subset of models are then discussed and analysed.


Rapid rotation makes the seismic interpretation of the oscillation
spectra of $\delta$ Scuti stars difficult, specially regarding the
mode identification. For very fast rotators, it has been shown
(Lignieres et al. 2006) that the rotation-pulsation interaction cannot
be described using a perturbation theory. The non-perturbative theory
for the oscillations computation applied to polytropes predict that
both small and large spacings are affected by rotation. Following
Reese et al. (2008) we estimate (within the range [250, 450] $\mu$Hz)
such an effect on the large separation is about 6 $\mu$Hz for the
models considered in this work.\footnote{Predictions obtained for a
  n=3 polytropic model.} This value remains within the uncertainties
in the determination of the large separation given in Sect. 2,
allowing us to use non-rotating models to restrict the space of valid
models as described in the previous section. This allows us to use
non-rotating models to restrict the space of valid models as described
in the previous section.

Equilibrium non-rotating models were computed in the manner described
in Sect. \ref{sec:models}. Only models showing large separations
within the range of [45 ,60] were considered.

For the $\delta$ Scuti stars, the range of the observed frequencies is
far from the asymptotic regime and a clear periodicity is not
expected. However, as it is demonstrated in
Sect. 2, a pattern, identifiable as a large
separation, can be indeed found in the data (Fig. \ref{fig:large-sep},
left panel).

We thus searched for similar regularities in the models (see
Fig. \ref{fig:large-sep}, right panel). We found that the peak at
26.61 $\mu$Hz (and sub-multiples) is the half of the probable large
spacing, allowing us to discard the models that have not reached the
observed large separation. Note that only modes with frequencies in the
range [100,700] and with $\ell=0-3$ were taken into account.
Additional constraints come from the stability analysis of the modes,
as it is illustrated in Fig. 3.

Following an iterative method, the original uncertainty box can be
reduced drastically (Fig. 4). The preliminary result obtained is that
we have the following ranges in age, mass and radius of HD\,174936:
age [788.5, 1705.9] Myr, mass [1.47, 1.82] $M_{\odot}$ and radius [1.61,
  2,05] $R_{\odot}$. This implies a drastic decrease of the mass
uncertainty from a 54\% (without asterosiemology) to a 23\% (with
asteroseismology). A rough estimate of the rotation effects on these
seismic models predicts a variation of 0.03 $M_{\odot}$ in mass, which
is lower than the accuracy here obtained.

Finally, using a representative model for this star, we have studied
if the star can have, from an energetic point of view, all these modes
exited at the same time, with the result that all these modes need a
very small amount of stellar energy (Moya \& Rodr\'iguez-L\'opez 2010).

\section{Conclusions \label{sec:conclusions}}

We have performed an asteroseismic analysis of the $\delta$ Scuti star
HD\,174936, observed by CoRoT during its first short run, SRc01. The
very precise space photometry provides the possibility of dealing with
a significantly large number of frequencies (around 400), for which we
demonstrated that the star has enough energy to excite. However,
Kallinger \& Matthews (2010) have recently suggested that most of the
lowest amplitude frequencies can be a consequence of surface
granulation.

We have combined the classical seismic analysis with the use of
statistical properties of the modes. In particular, we have searched
for periodic patterns in the frequency spectrum of HD\,174936 in order
to find new observational constraints. We have found a peak
distribution in the frequency spectrum that seems to correspond with a
large separation about 50~$\mu$Hz.

We have then performed a theoretical analysis in which models have
been constrained to fit the observed large separation and to predict
unstable the observed modes. These restrictions yield a range of
models with [7801, 8192]~K in $T_{eff}$ and [4.07, 4.19] in $\log g$,
which corresponds to a range of [1.47, 1.82]~$M_{\odot}$ in mass,
[788.5, 1705.9]~Myr in age and [1.61, 2,05]~$R_{\odot}$ in radius.

\acknowledgements {A.G.H. and R.G. acknowledge support from the
  Spanish ``Plan Nacional del Espacio" under project
  ESP2007-65480-C02-01. A.G.H. acknowledges support from a ``FPI"
  contract of the Spanish Ministry of Science and Innovation. CRL
  acknowledges an \'Angeles Alvari\~no contract under Xunta de
  Galicia.JCS acknowledges support from the "Instituto de
  Astrof\'{\i}sica de Andaluc\'{\i}a (CSIC)" by an "Excellence Project
  post-doctoral fellowship" financed by the Spanish "Conjerer\'{\i}a
  de Innovaci\'on, Ciencia y Empresa de la Junta de Andaluc\'{\i}a"
  under proyect "FQM4156-2008".}


\end{document}